\documentclass[12pt,preprint]{aastex}
\shorttitle{Velocity Dispersion Profiles of UMi and Dra}
\shortauthors{Mu\~noz et al.}

\begin{document}

\title{Exploring Halo Substructure with Giant Stars:
The Velocity Dispersion Profiles of the Ursa Minor and Draco 
Dwarf Spheroidals At Large Angular Separations}

\author{
Ricardo R. Mu\~noz\altaffilmark{1},
Peter M. Frinchaboy\altaffilmark{1},
Steven R. Majewski\altaffilmark{1},
Jeffrey R. Kuhn\altaffilmark{2},
Mei-Yin Chou\altaffilmark{1},
Christopher Palma\altaffilmark{3},
Sangmo Tony Sohn\altaffilmark{1,4},
Richard J. Patterson\altaffilmark{1} \&
Michael H. Siegel\altaffilmark{1,5}}

\altaffiltext{1}{Dept. of Astronomy, University of Virginia,
Charlottesville, VA 22903-0818 (rrm8f, pmf8b, srm4n, mc6ss, rjp0i@virginia.edu)}

\altaffiltext{2}{Institute for Astronomy, University of Hawaii, Honolulu HI 96822
(kuhn@ifa.hawaii.edu)}

\altaffiltext{3}{Dept. of Astronomy \& Astrophysics, Penn State, University Park, PA 16802
(cpalma@astro.psu.edu)}

\altaffiltext{4}{Korea Astronomy and Space Science Institute, 61-1 Hwaam-Dong,
Yuseong-Gu, Daejeon 305-348 Korea (tonysohn@kasi.re.kr)}

\altaffiltext{5}{University of Texas -- McDonald Observatory
Austin, TX  78712 (siegel@astro.as.utexas.edu)}

\begin{abstract}
  
  We analyze velocity dispersion profiles for the Draco and Ursa Minor
  (UMi) dwarf spheroidal (dSph) galaxies based on published
  and new Keck HIRES spectra for stars in the outer UMi field.
  Washington$+DDO51$ filter photometric catalogs provide additional
  leverage on membership of individual stars, and beyond 0.5 King
  limiting radii ($r_{\rm lim}$) identify {\it bona fide} dSph members up to 4.5
  times more efficiently than simple color-magnitude diagram selections.
  Previously reported ``cold populations'' at $r_{\rm lim}$ are not
  obvious in the data and appear only with particular binning; more or less constant
  and platykurtic dispersion profiles are characteristic of these dSphs
  to large radii.  We report discovery of UMi stars to at least
  $2.7r_{\rm lim}$ (i.e., $210\arcmin$ or 4 kpc).  Even with conservative
  assumptions, a UMi mass of $M>4.9\times10^8$ $M_{\sun}$ is
  required to bind these stars, implying an unlikely global
  mass-to-light ratio of $M/L>900$ $(M/L)_{\sun}$.  We conclude that we
  have found stars tidally stripped from UMi.
\end{abstract}


\section{Introduction}
Dwarf spheroidal (dSph) galaxies are 
thought to be strongly dark matter
(DM) dominated, with global mass-to-light $([M/L]_{\rm tot})$ ratios ranging
from a few to hundreds in solar units.  Spectroscopic databases are now
available for some dSphs to their King limiting radii, $r_{\rm lim}$
(\citealt{Kleyna2004}; \citealt[][hereafter {W04}]{W04}), and beyond
(\citealt{Tolstoy2004}; \citealt{Westfall2005}; R. Mu\~noz et al., in
preparation, hereafter M05),
enabling investigation of
the dSph kinematics and inferred mass distribution to large
radii.  Yet, very low stellar densities, 
still formidably challenge efficient spectroscopic study of
dSphs at large angular separations.  The photometric filtering 
techniques that are the basis of this
series of papers successfully overcome this problem and can
substantially increase the radial extent of dSph dynamical surveys.

Derived radial velocity (RV) dispersion ($\sigma_{\rm v}$) profiles for dSphs
tend to remain rather flat to well past the core radius.
\citet{Kleyna2002} attempted to fit the flat Draco (Dra) dSph profile
using two-parameter spherical models (\citealt{Wilkinson2002}) that
yield increasing $M/L$ with radius and a net $(M/L)_{\rm tot}$ of
$440\pm240$.  \citet{lokas2002}, applying a constant anisotropy
parameter model to the $\sigma_{\rm v}$ profiles of the Fornax and Dra dSphs,
derived $\sim10^9$ $M_{\sun}$ masses for these two systems.
Cosmology-dependent studies (\citealt{Stoehr2002};
\citealt{Hayashi2003}) based on $\Lambda$CDM models interpret the flat
$\sigma_{\rm v}$ profiles as consistent with massive DM halos surrounding
luminous cores. This interpretation helps alleviate the ``missing
satellites'' problem (\citealt{Kauffmann1993}; \citealt{Klypin1999};
\citealt{Moore1999}) endemic to these cosmologies.  
MW tidal effects on dSphs have also been considered
(e.g., \citealt{hm69}; \citealt{kuhn1989}; \citealt{kuhn1993}; \citealt{Kroupa1997}; 
\citealt*{GFM99}; \citealt{fleck2003}; M05) with predictions of potentially 
significant unbound stellar populations producing flat/rising dSph $\sigma_{\rm v}$ profiles.

\citeauthor*{W04} and \citet{Kleyna2004} recently reported flat
or slightly rising $\sigma_{\rm v}$ profiles that suddenly decline near the
$r_{\rm lim}$ of the Dra, Ursa Minor (UMi) and Sextans dSphs.  Such widely
separated cold populations in dSphs are of interest because they have
been interpreted as signs of {\it mild} tidal disruption, 
but more
importantly, because they severely mitigate against the idea of
extended, heated populations around dSphs (\citeauthor*{W04}).  
How far dSphs really
extend as well as the bound versus unbound nature of putative
``extratidal'' components remains unanswered.  Spectroscopic observation
of dSph-associated stars beyond $r_{\rm lim}$ is thus important to confirm
the reality of the extended populations and to ascertain their
dynamical state.

In this $Letter$ the \citeauthor*{W04} RVs up to $r_{\rm lim}$ for Dra and
UMi are combined with new RVs for UMi stars to several $r_{\rm lim}$ to
reassess the $\sigma_{\rm v}$ profiles of these dSphs.  Washington$+DDO51$ filter photometry
aids our discrimination of dSph giant star members.
We show that: (1) UMi members exist well past $r_{\rm lim}$ (\S3).
(2) The $\sigma_{\rm v}$'s of Dra and UMi remain more or less constant to past
$r_{\rm lim}$ (\S5).  (3) A cold population at $r_{\rm lim}$ for one of
the dSphs is found only under certain binning schemes,
and furthermore, depends on how one defines outliers (\S5).  
(4) The Washington+$DDO51$
method is at least 5 times more efficient at finding {\it bona fide}
dSph members than color magnitude diagram (CMD)-selection schemes (\S4).

\section{Photometric and Spectroscopic Data}

Previous contributions in this series (\citealt{Maj2000a};
\citealt{Maj2000b}; \citealt[][hereafter {P03}]{P03};
\citealt{Westfall2005}) show that Washington$+DDO51$ photometry
effectively identifies rare dSph giant star members against
the high Milky Way foreground in the low density wings of dSphs (see \S4).  
We use a similar methodology here, where giant star
candidates are first selected within the two-color diagram (2CD) boundary
shown in Fig.\ 1.  The UMi photometry is from \citeauthor*{P03}
supplemented with similar Mosaic camera data along the northeast UMi major axis taken with
the Mayall 4-meter telescope on UT 2002 May 4-6.  The Dra photometry was
obtained with the MiniMosaic (MiniMo) camera and WIYN telescope on UT
2004 March 12-13 and 2005 April 16-19.  MiniMo's field of view
is not optimal for large area photometric
surveys; these data were taken as a backup project when the instrument
for our primary observing program failed.  As a consequence we only
partially surveyed Dra, covering a total area of $\sim0.84$ deg$^{2}$.  
Median photometric errors for this dataset are 
$(\sigma_M,\sigma_{T_2},\sigma_{DDO51}) =$ (0.014, 0.014, 0.018) at $M=21.0$

We have been unable to obtain spectroscopic follow-up of Dra giant
candidates identified with our photometry.  However, M. Wilkinson
graciously provided the \citeauthor*{W04} RV database for 416 and 266
observed stars in the directions of the Dra and UMi dSphs,
respectively.  Typical RV errors for these data are 2.4 (Dra) and
$2.9\,{\rm km\,s^{-1}}$ (UMi).  We cross-identify by celestial
coordinates 254 UMi field stars (95\%) but only 212 Dra field stars
(51\%) between the W04 and our databases.

To these data we add Keck HIRES \citep{vogt1994} spectra for 52 UMi
field stars obtained UT 2002 May 21-22 and UT 2004 May 12-13 (data available
from authors).
We deliberately targeted UMi giant candidates at large radii.  The
spectra were reduced with standard IRAF echelle reduction methodology
with RVs determined using the {\ttfamily fxcor} package on the 
Mg triplet ($5130-5210\AA$) order.  Our quoted RV
errors are determined as in \citet{Vogt1995} with a median of 
$4.3\,{\rm km\,s^{-1}}$, calibrated by
19 multiple measures of six different UMi stars.
For nine stars in common with \citet*{a95} we obtain a mean RV 
difference of $0.3\,{\rm km\,s^{-1}}$ 
with dispersion of only $2.9\,{\rm km\,s^{-1}}$.  The union of our
data with that of W04 yields 309 unique RVs for the UMi field.

\section{Revisiting dSph Membership}

The Washington$+DDO51$ database can be used not only to select dSph
giant star candidates {\it before} spectroscopy (as done here for
UMi) but for after-the-fact assessment of likely membership of stars
in existing RV catalogues.  Figure 1 shows our CMD and 2CD
for UMi and Dra.  Filled/open circles designate RV stars more/less
likely to be giants based on positions in the 2CD (with ``giants'' 
adopted as stars bounded by the thin line). A few stars just outside
our giant selection criteria that have an RV consistent with that of the dSph
have been considered giants as well.  
The ``likely giant stars'' tend to lie closer to the red/asymptotic giant 
branches of the dSphs than the ``less likely giants'' (Fig.\ 1).

Figure 2 shows RVs for all stars versus elliptical radius, $r_{\rm e}$,
normalized to $r_{\rm lim}=77\farcm9$ for UMi (\citeauthor*{P03}) and
$40\farcm1$ for Dra (\citealt{Odenkirchen2001}).  An ``elliptical
radius'' corresponds to the semi-major radius of the ellipse centered on
the dSph that intersects the position of the star and has the measured
ellipticity of the dSph (0.54 for UMi and 0.29 for Dra; from above references).  
We adopt elliptical rather than circular radii to follow the distribution
of stars, although 
a dSph's gravitational potential and tidal boundary do not necessarily 
mimic its observed shape (we revisit circular radii in \S5).

The \citeauthor*{W04} $3 \sigma$ rejection criterion to discriminate
likely dSph members (dashed lines in Fig.\ 2) corresponds 
to $\pm39\,{\rm km\,s^{-1}}$ around $<RV>=-290.8\,{\rm km\,s^{-1}}$ for Dra and
$\pm36\,{\rm km\,s^{-1}}$ around $<RV>=-245.2\,{\rm km\,s^{-1}}$
for UMi.  By these criteria (Fig.\ 2)
only six of our photometric Dra giant candidates lie outside this velocity range,
proving the reliability of our photometric discrimination technique.  
Among stars classified as giants by our photometric technique, two 
(solid squares in Fig.\ 2)
lie just outside the RV range at $-246.1\pm4.6\,{\rm km\,s^{-1}}$ and
$-332.03\pm4.8\,{\rm km\,s^{-1}}$ (uncertainties within
$1\sigma$ of the ``Dra member" RV limit).  While for $\sim200$ Dra members one
expects only $\sim0.5$ outliers at $>3\sigma$ for a {\it Gaussian}
distribution, the kurtosis excesses ($\gamma_2$) of {\it both} the Dra
and UMi RV distributions flatten from near Gaussian ($\gamma_2=0$) to
$\gamma_2=-0.8$ and $-0.9$, respectively, at $r_{\rm e} > 0.4r_{\rm lim}$, so
that more {\it apparent} ``outliers'' might be expected.
Based on these two arguments and the ``giant star" colors of these two stars
we consider them to be very likely Dra members.

In UMi, a larger number of giant candidates than in Dra live clearly
outside the RV criteria (Fig.\ 2).  Because even the faintest stars with
RVs have quite small photometric uncertainties (\citeauthor*{P03}), it
is not likely that these ``false positives'' are due to photometric
error, but rather represent field halo giants or metal-poor dwarf
stars with weak Mgb+MgH absorption.  Among the ``non-UMi'' giant candidates 
there appears to be RV
clumpiness, with 9/9/7 stars having
$\sigma_{\rm v}=9.2\pm2.3$/$9.6\pm2.6$/$16.6\pm4.6\,{\rm km\,s^{-1}}$ around
$<RV>=6.4/-54.8/-163.8\,{\rm km\,s^{-1}}$, respectively.  These RV
clumps of giant candidates have $\sigma_{\rm v}$'s of order those observed
in UMi and Dra, as well as
in the Sagittarius (Sgr) tidal tails (\citealt{SgrII}) and are
reminiscent of foreground halo substructure discovered in our similar survey
of Carina (M05); our giant star identification method may be finding
{\it other} halo substructure in the UMi field.  Nevertheless, the
majority of stars we photometrically classify as giants do lie inside
the \citeauthor*{W04} ``UMi'' RV range (see \S4).  A remaining
photometric giant candidate ({\it triangle}) lies just outside the UMi
RV limits but well inside the giant region in the 2CD {\it and} along
the RGB locus for UMi; we strongly suspect this star is a UMi member and
include it in our $\sigma_{\rm v}$ analysis (\S5).  In the end, 182 UMi and
210 Dra stars are included in our $\sigma_{\rm v}$ profiles (\S5).

\section{Photometric efficiency}

Particularly at large angular radii, the sky density of dSph stars with
brightnesses amenable to current spectroscopic capability (i.e., red
giants) is swamped by foreground stars.  To improve overall efficiency
of target selection, various groups (e.g., \citeauthor*{W04}) select
dSph targets by position on/near the dSph's giant branch in the CMD.
This typically decreases foreground contamination by an order of
magnitude but ``false positive'' sources still well outnumber dSph stars
outside $r_{\rm lim}$ (\citealt{MJ05}; \citealt{Westfall2005}).  However,
our Washington$+DDO51$ photometric technique improves sample reliability
by an additional order of magnitude over, for example, the
\citeauthor*{W04} CMD selection.

Over all angular separations (to $r_{\rm e} = 7.7r_{\rm lim}$ in the case of UMi)
our technique yields a dSph member identification efficiency of 87\% for
UMi and 97\% for Dra.  Our ``false positive'' identifications all lie
beyond $0.5r_{\rm lim}$ \citep[we ignore the possibility that their RV
clumping is due to wrapped UMi tidal debris arms, as observed in the Sgr
system;][]{MSWO}.  \citeauthor*{W04}'s overall success rate for their
CMD-selected candidates, to only $r_{\rm e}\sim 0.8r_{\rm lim}$ for UMi and
$r_{\rm e}\sim 1.8r_{\rm lim}$ for Dra, is 62\% and 50\%, respectively.
Considering targets with $r_{\rm e}/r_{\rm lim}>0.5$, the W04 efficiency drops
to 38\% (28/73) and 19\% (45/239) respectively, whereas our selection
method at similar radii still nets an overall efficiency of 77\%
(23/30) and 85\% (23/27) for UMi and Dra.  This 2 to 4.5 times greater member selection
efficiency optimizes the exploration of 
low density dSphs galaxies with valuable 6 to 10-m class telescope time.

\section{Velocity Dispersion Profiles and Interpretation}

Our $\sigma_{\rm v}$ profiles are computed using equal sample-size binning,
but similar results are found with equal, linear bin sizes (though some
bins are poorly populated in this scheme).  
The method of \citet{a86} is used to calculate $\sigma_{\rm v}$, 
because a Maximum Likelihood method like that used by \citeauthor*{W04} 
assumes a Gaussian velocity distribution at all radii, which is not observed for either 
dSph, and is also not expected in a 
disrupted system (M05).
To assess the influence of bin size and geometry Figure 3 
shows the UMi dispersion profile versus
elliptical (panel a) and circular radii (panel b) with 17, 12
and 7 members per bin (and the outermost bins accumulating any odd,
extra star).  As expected, profile variability is less pronounced as the
number of stars per bin increases.  The general UMi trend is an initial
decline in $\sigma_{\rm v}$ followed by a gentle rise and then a slightly
decreasing profile.  
The sudden decline to a ``cold point'' reported by
\citeauthor*{W04} appears only in the highest resolution binning that is
most susceptible to statistical fluctuations.  
While larger samples of stars in the outer regions of dSphs would
be helpful, UMi seems to share the trend of more or less flat 
dispersion profiles observed in the outskirts of other
dSphs (\citealt{Mateo1997}, \citealt{Westfall2005}, M05).

The Dra profiles (Fig.\ 4) use 21, 16 and 8 stars per bin, 
with solid circles for the 208 star samples.  Open symbols
show the outer profiles when the stars marked as squares in Figure 2
are included.  Except for a single bin in the lower profile of panel (b),
the same general $\sigma_{\rm v}$ behavior is seen at large radii as in UMi: a
flat (or maybe slowly decreasing) profile (depending on the binning) is
observed past $r_{\rm lim}$. In fact, 
a non-binning test, like the ``one-by-one'' test used
by \citet{Kleyna2004} for Sextans, does not show a cold population in
the outermost dispersion points for either dSph.
Moreover, a $\chi^2$
fit of the $\sigma_{\rm v}$ distribution to a constant value for both UMi and Dra shows
that there is no evidence that the outermost stars have a $\sigma_{\rm v}$ that is statistically
different from those of stars at smaller projected radii.

Tidal disruption can create an unbound stellar population near a dSph.
This seems to be a natural explanation for producing
{\it both} the observed extended stellar distributions {\it and} 
flat/slowly declining velocity
dispersion profiles at large radii in dSphs (\citealt{kuhn1989}; \citealt{Kroupa1997}; 
\citealt{Mayer2002}, \citealt{fleck2003}; M05). 
While the flat $\sigma_{\rm v}$ profiles observed in UMi and Dra have
also been modeled by an {\it ad hoc} extended DM halo (\citealt{Kleyna2002}), 
two additional piece of evidence support a tidal disruption scenario.  First,
the observed platykurtic velocity distributions at large radii in both UMi and
Dra more closely match the flattened RV distributions of unbound dSph stars at
large radii in detailed N-body tidal disruption simulations (e.g. M05).

Second, our new Keck RVs have verified the most widely separated
member stars for any dSph other than the tidally disrupting Sgr galaxy.
Our most separated UMi-field star having a UMi RV is at a linear distance of $238\arcmin$
(4.8 kpc) and near the minor axis; this star's elliptical radius of
$r_{\rm e} = 6.6r_{\rm lim}$ implies a major axis radius of 10.4 kpc if the UMi
ellipticity is maintained at all radii.  Assuming a spherical
potential for UMi, the required mass to keep this star bound to the
dSph is $M > 3 M_{\rm MW} (r/R_{\rm GC})^3$, or $7.6\times10^8$ $M_{\sun}$
for a Milky Way mass within the UMi distance ($R_{\rm GC}=69$ kpc) of
$M_{\rm MW}=7.6\times10^{11}$ $M_{\sun}$ (\citealt{Burkert1997}); this translates to
$(M/L)_{\rm tot} > 1,400 $ $(M/L)_{\sun}$ when we adopt $L=5.4\times10^{5}
L_{\sun}$ (\citeauthor*{P03}).  However, if the tidal boundary of UMi is more
elongated, a larger $M/L$ is implied; at the limit where the tidal boudary is elongated
according to its central ellipticity, this star, were it placed at its
corresponding major axis radius (i.e., $r=r_{\rm e}$ or 10.4 kpc), implies an
astounding UMi $(M/L)_{\rm tot} > 14,400$ $(M/L)_{\sun}$ for UMi.

It might be argued that the latter star is a field halo interloper, since its RV
is only $\sim60$ km s$^{-1}$ (to the retrograde side) of that expected
for a non-rotating halo (assuming a 232 km s$^{-1}$ solar rotation about the 
Galactic center) and within
the $\sigma_v$ of a hot MW halo ($\sim100$ km s$^{-1}$; e.g., \citealt{sirko04}).
Higher $S/N$ spectroscopy to test this star's chemical properties would be useful.
However, under the above MW dynamics,
the second most separated ``RV member" --- 
at $210\arcmin$ (4 kpc), or $r_{\rm e} = 2.7r_{\rm lim}$ along the UMi major axis ---
is retrograde by $\sim105$ km $^{-1}$.
A UMi-bound star at this projected radius implies a UMi mass $>4.9\times10^8$
$M_{\sun}$ or $(M/L)_{\rm tot} >900$ $(M/L)_{\sun}$.

One could use a Milky Way model, 
like the Besancon model ($http://bison.obs-besancon.fr/modele/$)
to estimate the expected Milky Way background; 
this model predicts $\sim 3.2$ MW halo giants
in the range of RV, color and magnitude of the present 9.7 deg$^2$ UMi survey 
area outside $r_{lim}$, under the assumption that we have observed {\it all} possible targets 
in the field, whereas we have only observed 37\% (implying about 1 interloper in our $>r_{lim}$ RV sample). 
However, the Besancon model uses smooth population density laws to describe
{\it mean} densities and does not account for second order substructure perturbations.
In the case that our color-color-magnitude-RV sampling of parameter space
happens to be dominated by a substructure ``void", the Besancon model background
should be an overestimate.  Were our sampling instead {\it overpopulated} by substructure, 
the model would underestimate the background.  But the presence of the substructure
should be obvious by coherence in our parameter space, while, for there to be 
interlopers within our UMi sample, these stars would have to both lie at roughly the UMi distance
(to share position in the CMD and 2CD) and have about the UMi RV --- a situation
we consider improbable.  Moreover, the Besancon model in fact {\it overpredicts}
by $2.3\times$ the number of Milky Way stars we see {\it outside} the UMi RV-member 
range; accounting for this factor, in the limit of a smooth halo, 
implies only 0.5 Milky Way contaminants within the UMi RV sample.  In the end, we
regard as unlikely that {\it both} of the outermost UMi RV members are
contaminants.

If either of these outer dSph stars are bound, UMi has the largest $M/L$ of any
galaxy by yet {\it another} order of magnitude {\it or two} than previously suggested.
The corresponding physical dimensions
of UMi would rival the King profile extent of the Sgr density
distribution, of which a significant fraction, however, has been shown must
be {\it unbound} (\citealt{MSWO}).  From the extreme physical dimensions
implied it is difficult to avoid the conclusion that true extratidal
stars have now been identified for the UMi system ---
results consistent with suggestions of tidal 
disruption of this dSph by
photometric analyses as early as that of \citet{hodge} and more recently
by \citet{Delgado2001}, \citeauthor*{P03}, and
\citet{Gomez-Flechoso2003}.
The existence of shared photometric and RV trends
between UMi and Dra to at least $r_{\rm lim}$ points to a possible
tidal disruption scenario for Dra as well \citep*[e.g.,][]{smith97}.

We gratefully acknowledge support by NSF grant AST-0307851, NASA/JPL
contract 1228235, the David and Lucile Packard Foundation, Frank
Levinson through the Celerity Foundation, the Virginia Space Grant 
Consortium, and the IfA/UH.  We thank the referee, N. Wyn Evans for
helpful suggestions to improve the paper.

\begin{figure}
\plotone{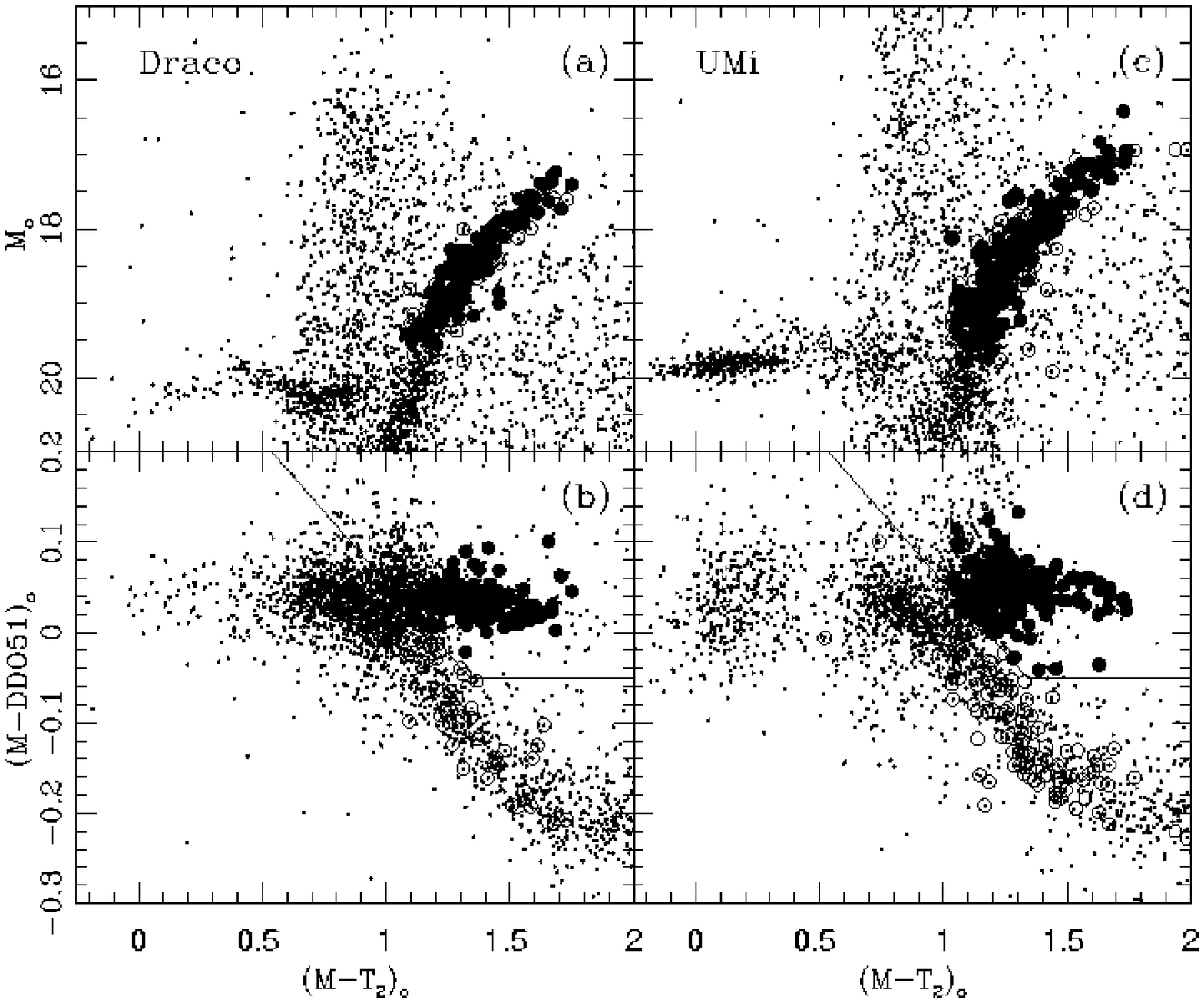}
\caption{(a)
  Color-magnitude diagram (CMD) for the Dra dSph photometric catalog.
  Solid/open circles show stars with available RV data selected/not
  selected photometrically to be Dra giants based on panel (b).  (b) $(M
  - T_2, M - DDO51,)_o$ diagram for the same data as panel (a). (c) and
  (d): CMD and 2CD for the UMi dSph from \citeauthor*{P03}.  For
  clarity, only stars within one $r_{\rm lim}$ have been plotted.  Symbols
  in (c) and (d) have similar meaning as for panels (a) and (b) but for
  the UMi field.  }
\end{figure}

\begin{figure}
\plotone{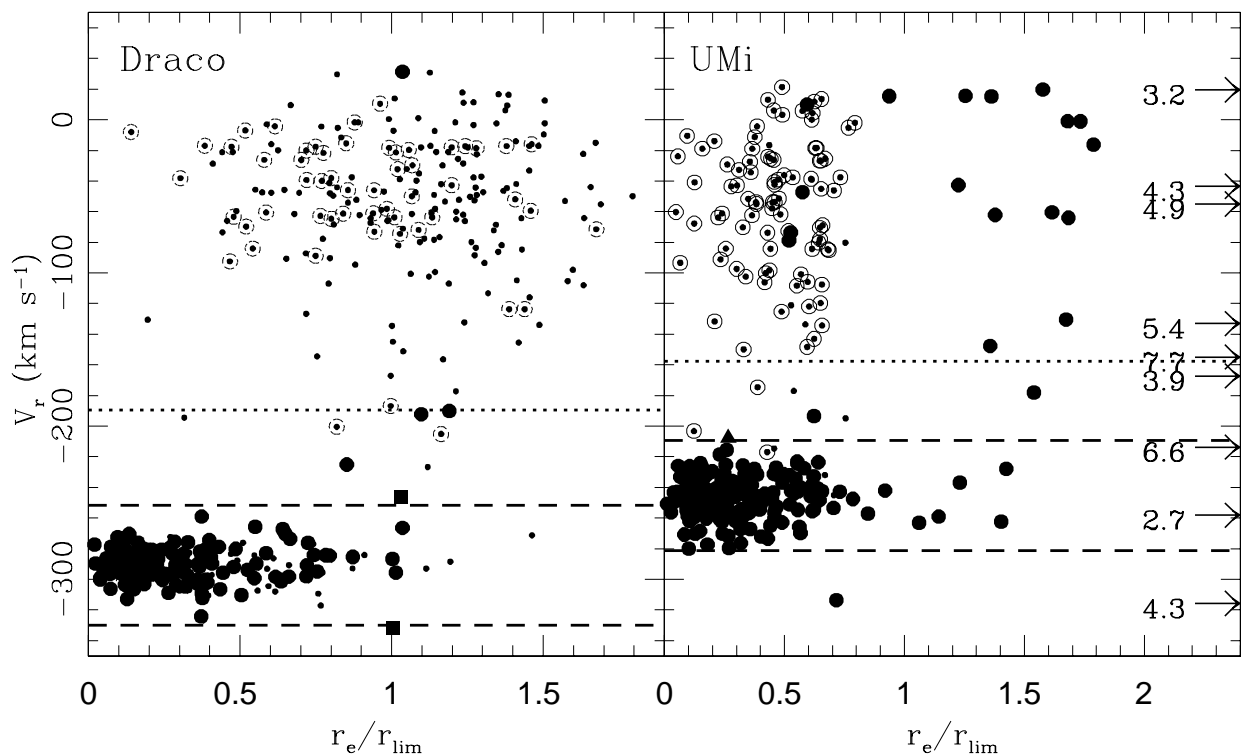}
\caption{RV versus elliptical radius (normalized to $r_{\rm lim}$)
  for the Dra (a) and UMi (b) dSphs.  Symbols are as in
  Fig.\ 1. Dashed lines delineate the $3\sigma$ RV range adopted as dSph
  membership criteria by W04.  Arrows indicate the RVs of stars outside
  the plotted area (normalized radii indicated for these stars next to their
  arrows).  Dotted lines show RV expected for zero Galactic
  rotation velocity, assuming a 232 km s$^{-1}$ solar rotation velocity about the MW.}
\end{figure}

\begin{figure}
\plotone{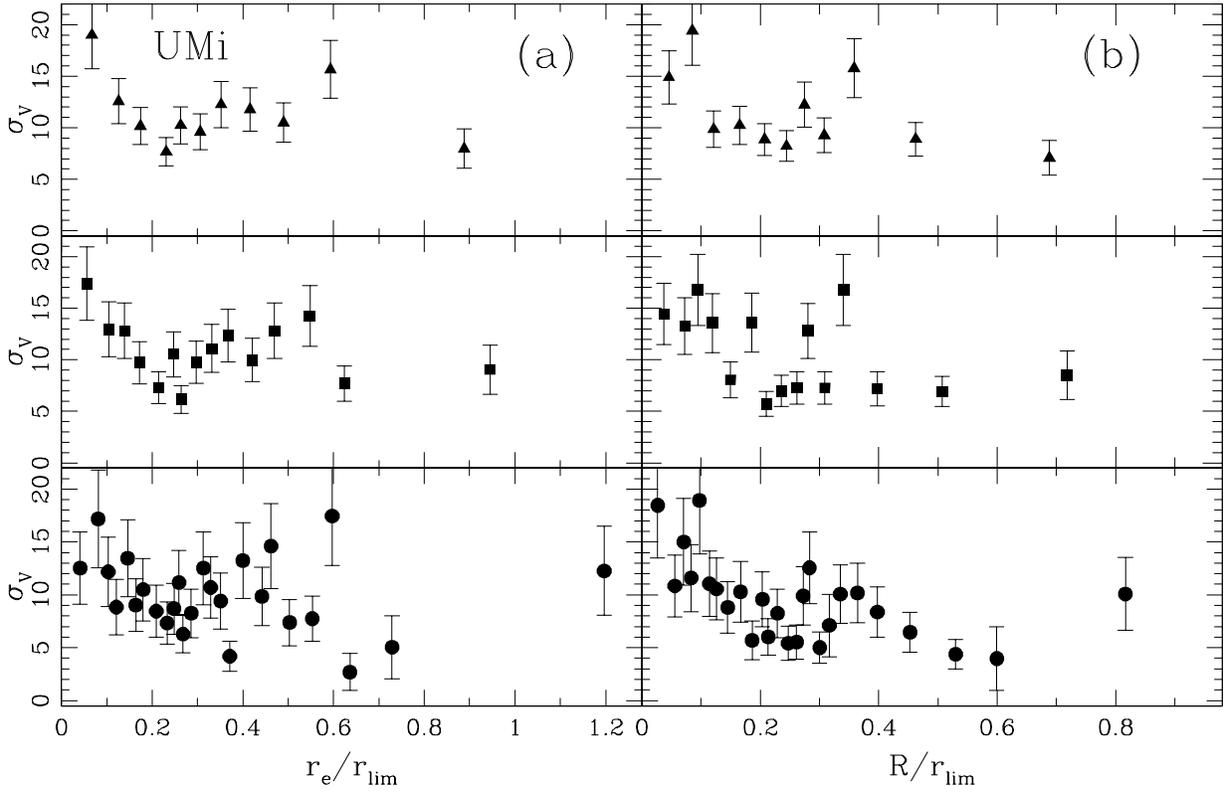}
\caption{RV dispersion profiles for the UMi dSph.
  The panels use (top to bottom) 17, 12 and 7 stars per bin.  Solid symbols 
  show $\sigma_{\rm v}$'s calculated from the 182
  star sample. Panels show profiles versus (a) elliptical and (b)
  circular radii.}
\end{figure}

\begin{figure}
\plotone{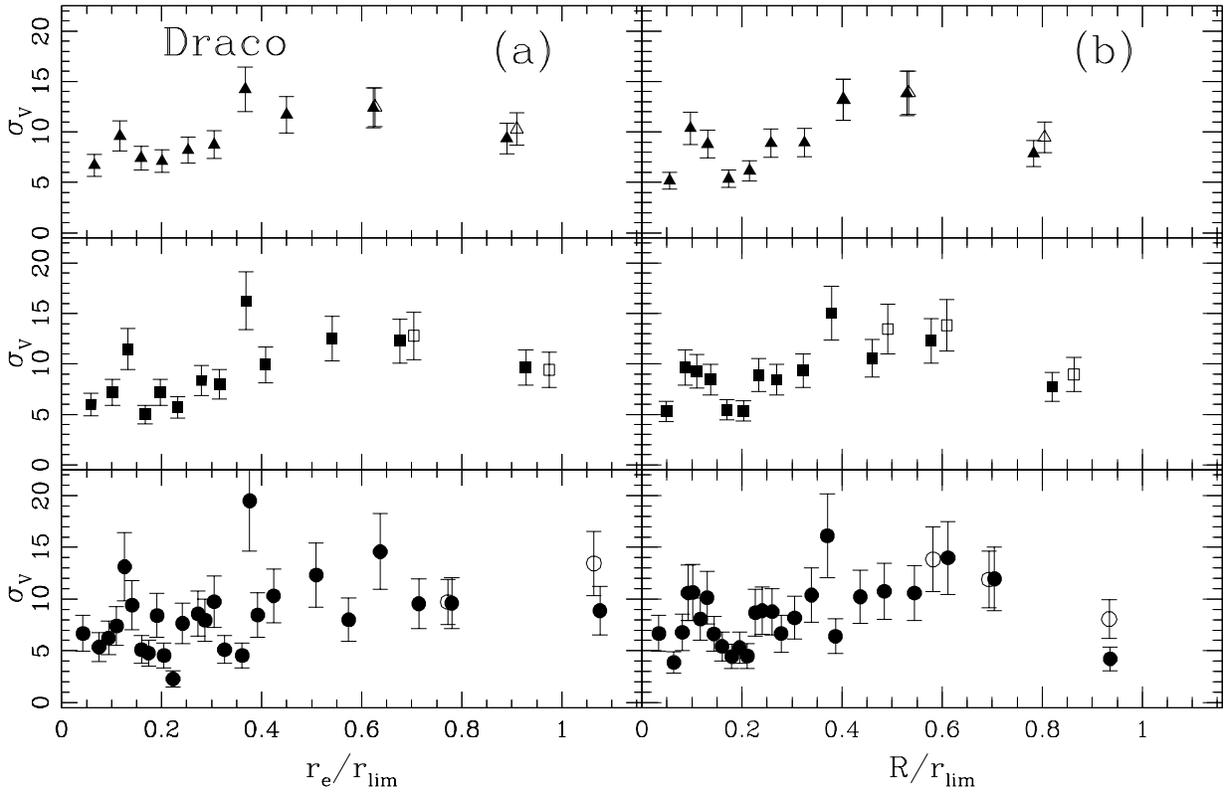}
\caption{Same as Fig.\ 3 for the Dra dSph.
  From upper to lower panel, 21, 16 and 8 stars are used per bin.  Open 
  symbols show the profile when the stars shown by
  squares in Fig.\ 2 are included.}
\end{figure}

\end{document}